\documentclass[preprint,12pt]{elsarticle}

\usepackage{amssymb}

\usepackage{amsmath,amsthm,amsfonts,bm}
\usepackage{subcaption}
\usepackage{hyperref}
\usepackage{stackengine}
\usepackage{parskip}
\usepackage{relsize}
\usepackage{tabularx}
\usepackage{booktabs}
\usepackage{soul}
\usepackage{algorithmic}
\usepackage[linesnumbered,ruled,vlined]{algorithm2e}
\usepackage{placeins}
\usepackage{float}
\usepackage{xcolor}
\usepackage[capitalise]{cleveref}

\newcommand{\ecgtizer}{\textsc{ECGtizer}}
\newcommand{\ecgrecover}{\textsc{ECGrecover}}
\newcommand{\ecgminer}{\textsc{ECGminer}}
\newcommand{\paperecg}{\textsc{PaperECG}}
\newcolumntype{L}[1]{>{\raggedright\arraybackslash}p{#1}}
\newcolumntype{C}[1]{>{\centering\arraybackslash}p{#1}}

\journal{Computer Methods and Programs in Biomedicine}

\begin{document}

\begin{frontmatter}



\title{\ecgtizer: a fully automated digitizing and signal recovery pipeline for electrocardiograms}

\author[1]{Alex Lence}
\ead{alex.lence@ird.fr}
\author[1]{Ahmad Fall}
\author[1,2]{Samuel David Cohen}
\author[1]{Federica Granese}
\author[1,3]{Jean-Daniel Zucker}
\author[2,4]{Joe-Elie Salem}
\author[1,3]{Edi Prifti}
\ead{edi.prifti@ird.fr}

\address[1]{IRD, Sorbonne Université, Unité de Modélisation Mathématique et Informatique des Systèmes Complexes, UMMISCO, F-93143 Bondy, France}

\address[2]{Clinical Investigation Center Paris-Est, CIC-1901, INSERM, Department of Pharmacology, Pitié-Salpêtrière University Hospital, Sorbonne Université, France}

\address[3]{Sorbonne Université, INSERM, Nutrition et Obesities; systemic approaches, NutriOmique, AP-HP, Hôpital Pitié-Salpêtrière, France}

\address[4]{Department of Medicine, Vanderbilt University Medical Center, Nashville, TN, USA}

\cortext[cor1]{Corresponding author}

\sloppy

\begin{abstract}
Electrocardiograms (ECGs) are essential for diagnosing cardiac pathologies, yet traditional paper-based ECG storage poses significant challenges for automated analysis. This study introduces {\ecgtizer}, an open-source, fully automated tool designed to digitize paper ECGs and recover signals lost during storage, facilitating automated analyses with modern AI methods. {\ecgtizer} employs automated lead detection, three pixel-based signal extraction algorithms, and a deep learning-based signal reconstruction module. We evaluated {\ecgtizer} on two datasets: a real-life cohort from the COVID-19 pandemic (JOCOVID) and a publicly available dataset (PTB-XL). Performance was compared with two existing methods: the fully automated {\ecgminer} and the semi-automated {\paperecg}, which requires human intervention. {\ecgtizer}’s digitization performance was assessed in terms of signal recovery and the fidelity of clinically relevant feature measurement. Additionally, we tested these tools on a third dataset (GENEREPOL) for downstream AI tasks. Results show that {\ecgtizer} outperforms existing tools, with its {\ecgtizer}$_{\text{Frag}}$ algorithm delivering superior signal recovery performance. While {\paperecg} demonstrated better outcomes than {\ecgminer}, it also required human input. {\ecgtizer} enhances the usability of historical ECG data and supports advanced AI-based diagnostic methods, making it a valuable addition to the field of AI in ECG analysis.
\end{abstract}

\begin{highlights}
\item Introduced ECGtizer, a fully automated tool to digitize paper ECGs.
\item ECGtizer outperforms current methods in signal recovery.
\item Tested on real-world and public datasets.
\item Supports advanced AI diagnostic methods.
\item Opens new research avenues by allowing the use of ECG historical data.
\end{highlights}


\end{frontmatter}


\section{Introduction}

An electrocardiogram (ECG) is a diagnostic tool that captures the heart's electrical activity by placing electrodes on the patient's body~\cite{Einthoven1903}. The resulting ECG waveform contains vital information, with minor variations potentially indicating a range of diseases. Historically, ECGs have been printed and stored on paper or saved as images, with typically only one quarter of the original signal preserved to save space and enhance readability. This approach presents clear long-term challenges for accessing and analyzing ECG data, not only in terms of accessibility, which demands significant human resources, but also its unsuitability for automated analysis, especially with novel AI-based methods.

In addition to providing ECG images in PDF files, ECG system manufacturers may also offer raw data, often encoded in proprietary formats. However, these raw files are generally lost due to the difficulty of long-term reliable storage in patients' medical records~\cite{santamonica2024ecgminer}. Many countries, particularly in developing regions, still rely on paper-based ECGs~\cite{Adedinsewo2022,Tun2017}. Numerous healthcare centers worldwide have amassed valuable cohorts of well-characterized patients with rare diseases, with ECGs recorded on paper over decades. However, accessing and analyzing these ECGs with specialized software remains challenging, which poses a significant barrier to clinical research and diagnosis advancement.

Over the years, several tools have been developed to convert ECG scans or PDF versions into numerical waveform signals~\cite{lence_automatic_2023}. Despite these advancements, most tools are not easily accessible~\cite{badilini_ecgscan_2005, randazzo_development_2022, baydoun_high_2019} and require human intervention, making large-scale digitization of ECG data challenging. Some newer tools have enhanced automatic ECG digitization capabilities~\cite{santamonica2024ecgminer, himmelreich2023diagnostic}, but they can only digitize the portion of the signal preserved in the ECG image and cannot recover the missing three-quarters of the signal lost during printing or PDF conversion.

Digitization serves two primary objectives. Firstly, it offers an alternative method for storing ECG records, mitigating issues like paper degradation and ink evaporation, while significantly reducing storage space~\cite{rajiv_gandhi_institute_of_technology_mumbai_india_robust_2017, chebil_novel_2008}. Secondly, it enables automatic signal analysis using AI-based systems, which are revolutionizing ECG interpretation~\cite{somani2021deep, wu_fully-automated_2022}. Training deep learning models on digital ECG data in numerical format proves more effective than using images, as it preserves the temporal sequence of the signal and simplifies preprocessing tasks such as normalization and pseudonymization, thereby enhancing model performance~\cite{mishra_ecg_2021}.

Research on ECG digitization has been ongoing since the 1990s, with a notable surge in interest since the mid-2000s~\cite{widman_digitization_1991, lence_automatic_2023}. This article focuses on the digitization of an ECG as the extraction of an equivalent numerical vector representation from an analog-recorded ECG signal image.

Here, we propose {\ecgtizer}, a fully automated open-source software that not only digitizes ECG images with high accuracy but also recovers the missing signal discarded during the printing process. {\ecgtizer} uses automatic lead detection through pixel variance estimation before extracting the signal pixel by pixel, adjusting it for both temporal and spatial scales (amplitude). The raw digitized signal is then input into the {\ecgrecover} module. Based on a deep learning UNet architecture~\cite{lence2024ecgrecoverdeeplearningapproach}, {\ecgrecover} uses the signal of other leads to regenerate the missing portions of the signal. The final output is a data frame containing the complete signal, which can be stored as a standalone XML file, and includes automated anonymization features to facilitate data sharing.

Specifically, our contributions are threefold:
\begin{enumerate}
    \item An open-source and fully automated software called {\ecgtizer} is provided  \footnote{Code available at: \href{https://github.com/UMMISCO/ecgtizer.git}{https://github.com/UMMISCO/ecgtizer.git}}.
    \item Three distinct lead-extraction algorithms — Full, Fragmented, and Lazy — were proposed and thoroughly evaluated in {\ecgtizer}. Performance compared with two state-of-the-art methods, {\ecgminer}~\cite{santamonica2024ecgminer} and {\paperecg}~\cite{Fortune2022} are given, focusing on their impact on downstream AI tasks like TdP-risk prediction. {\ecgtizer} outperformed these methods.
    \item The{\ecgrecover}~\cite{lence2024ecgrecoverdeeplearningapproach} module has been integrated into {\ecgtizer} for reconstructing complete ECG signals from incomplete parts, enhancing the accuracy and utility of downstream AI-based diagnostic tools.
\end{enumerate}

The article is organized as follows. In~\cref{sec:related}, we review related works. In~\cref{sec:methodology}, we detail the {\ecgtizer} approach, specifying the input and output at each stage and providing the pseudo-code for the lead-extraction algorithms. The performance of the digitization tools, both before and after the lead-completion process, is evaluated in~\cref{sec:results1}. The impact of digitization on downstream AI tasks, such as TdP-risk prediction, is analyzed in~\cref{sec:tdp}. Finally, we present the conclusion of the paper in~\cref{sec:conclusion}.


\section{Related Works}
\label{sec:related}
\subsection{Digitization Tools}

ECG digitization has emerged as a widely explored area of research. A recent review from 2023~\cite{lence_automatic_2023} identified and analyzed 63 pertinent articles, categorizing digitization tools into manual and fully automated types. Manual tools, such as those described in \cite{badilini_ecgscan_2005, Fortune2022, ravichandran2013novel}, require user interaction for tasks like lead delineation and threshold setting. Semi-automatic methods, exemplified by \cite{mallawaarachchi2014toolkit} and \cite{baydoun_high_2019}, also involve user intervention, typically for corrections during signal extraction. 

Fully automated tools, on the other hand, allow users to input scanned ECGs and automatically complete the digitization process, as seen with {\ecgminer} ~\cite{santamonica2024ecgminer} and our proposed software, {\ecgtizer}. Among the manual tools, we highlight {\paperecg} by Fortune et al.~\cite{Fortune2022}, which requires initial manual adjustments, such as image tilting and lead positioning, before proceeding to digitally extract each lead. In contrast, our {\ecgtizer} provides a fully automated solution for digitizing paper-based ECGs, simplifying the process and reducing the potential for human error.

\subsection{Digitization Process}
Overall, the digitization process typically involves three stages:
(i) \textit{lead detection}, (ii) \textit{image thresholding}, and (iii) \textit{lead extraction}.

\textbf{Lead detection}. 
An ECG is a stereotyped image; for the same format, the position of the leads is generally consistent. Lead detection involves segmenting the ECG image to localize each trace produced by the ECG leads. The leads typically appear as black traces on a colored or black-and-white grid. Their detection may require human intervention \cite{kumar_extracting_2012, mishra_ecg_2021, randazzo_development_2022}. Baydoun et al.~\cite{baydoun_high_2019} proposed an automatic lead detection approach that analyzes the variance in the image to distinguish between areas of significant variation and those with minimal variation (see~\cref{sec:trace_extraction}). In the vicinity of the lead line, pixels show significant peaks and variations that help pinpoint the positions of the ECG leads (~\cref{fig:CutImage}).

\textbf{Image thresholding}.
In most paper ECGs, the signal is traced in black on a red grid and white background. Removing the grid is crucial as it interferes with the signal extraction process. Image thresholding involves selecting an optimal threshold value that separates pixel intensities. Pixels above this threshold are classified as foreground (active), while those below are classified as background (inactive). This method effectively helps distinguish areas of interest from the rest of the image. Two thresholding approaches are commonly recommended: the Otsu method~\cite{otsu_threshold_1979} for sharp, high-quality images, and the Sauvola method~\cite{sauvola_adaptive_2000} for noisier images. More recent approaches include deep learning techniques~\cite{li_deep_2020}.

\textbf{Lead extraction}.  
This stage involves associating each pixel in the signal with an amplitude value and a corresponding time position. Methods for extracting leads include active contour approaches~\cite{badilini_ecgscan_2005}, which dynamically adjust contours around lead traces, and waveform path methods, which isolate and traverse each lead to extract individual pixels~\cite{wu_fully-automated_2022, santamonica2024ecgminer}.

Current literature on the digitization of paper ECGs highlights three main limitations: (i) automation of the process, (ii) access to the software's code, and (iii) the lack of benchmark datasets in original studies. To address these issues, we have identified two approaches that provide code access for direct comparison: \ecgminer~\cite{santamonica2024ecgminer}, a fully automated digitizing method, and \paperecg~\cite{Fortune2022}, which offers an API for digitizing ECGs but requires manual lead selection.

\section{Methods}
\label{sec:methodology}

We propose {\ecgtizer}, a fully automated software for ECG digitization. To address the critical problem of partial signal representation in ECGs, {\ecgtizer} includes a module named {\ecgrecover}~\cite{lence2024ecgrecoverdeeplearningapproach}, which utilizes a deep learning model based on the U-Net architecture to reconstruct the missing portions of the signal from existing segments of the 12 leads across a 10-second recording.

\subsection{The {\ecgtizer} Pipeline}
Below, we outline the {\ecgtizer} pipeline, which consists of five main steps: (i) conversion of the PDF to an image, (ii) removal of textual elements, (iii) identification of the traces, (iv) extraction of the signal, and (v) completion of the signal. 

\subsubsection{PDF to RGB-Matrix Conversion}
\label{sec:conversion}
The first step involves converting the \texttt{.pdf} file into an RGB matrix. This conversion is crucial, as it influences the quality of the subsequent digitization steps. We recommend a resolution of 300 dots per inch (DPI), which offers a balance between file size and detail preservation, crucial for maintaining the integrity of the ECG signals. However, {\ecgtizer} provides flexibility, allowing users to select their preferred resolution based on the specific requirements of their ECG data. This adaptability ensures that {\ecgtizer} can handle both vector-based images, typically generated during ECG acquisition, and scanned images, where resolution settings can significantly impact the quality of the digitized output.

\noindent\fbox{\begin{minipage}{\columnwidth}
\textbf{Input:} \textit{Paper-based ECG in \texttt{.pdf}, \texttt{.jpeg} or \texttt{.png} format.}\\
\textbf{Output:} \textit{A tensor containing the image encoded in RGB (red, green, blue), as three matrixes of size n$\times$m (the dimensions of the image in pixels).}
\end{minipage}}

\subsubsection{Text Removal}
\label{sec:text_removal}

Paper-based ECGs often contain metadata, including printed text, which provides automatic annotations of the ECG visit, and sometimes even manual manuscript annotations along with patient information. The challenge is to detect and remove this text from the image while preserving the electrocardiographic signal intact. For this purpose, {\ecgtizer} employs the \texttt{Pytesseract} library, an optical character recognition (OCR) tool\footnote{\url{https://github.com/madmaze/pytesseract}}. This OCR capability is critical for distinguishing and eliminating text without degrading the underlying ECG signal. If text removal is unsuccessful, it typically indicates that the text is superimposed directly on the signal, posing significant challenges for signal extraction without quality degradation.

\noindent\fbox{\begin{minipage}{\columnwidth}
\textbf{Input:} \textit{Paper-based ECG converted into a RGB matrix.}\\
\textbf{Output:} \textit{The RGB matrix, with the text pixels replaced by the average value of the surrounding pixels. The textual metadata is stored in a separate vector.}
\end{minipage}}

\subsubsection{Trace Detection}
\label{sec:trace_extraction}
\begin{figure*}[htbp] 
    \centering
    \FloatBarrier
    \includegraphics[width=1.\textwidth]{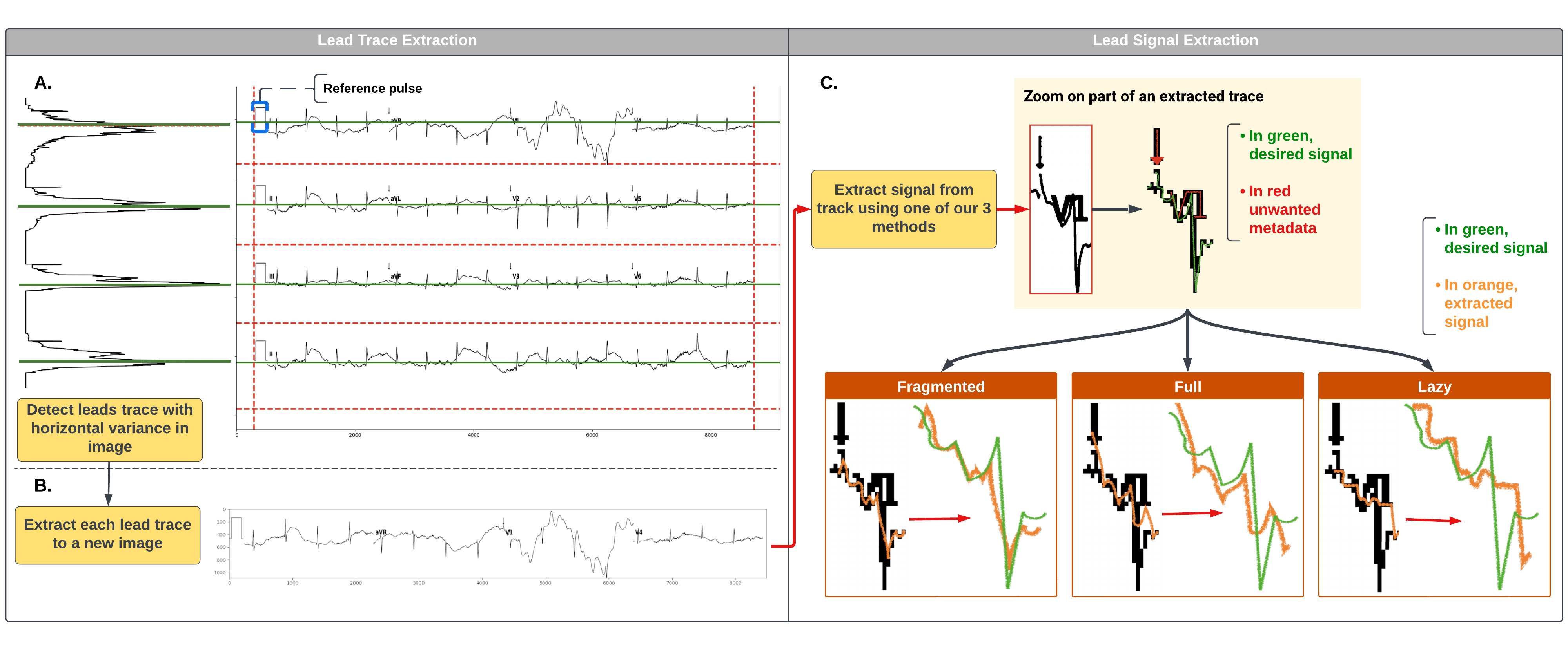}
    \caption{From left to right: In the first stage, \textit{Lead Trace Extraction}, the ECG image is binarized using the Otsu thresholding method. Pixel variance is then calculated to assess intensity variation across columns and rows, enabling traces extraction. In the second stage, \textit{Lead Signal Extraction}, {\ecgtizer} analyzes each column of the extracted trace image matrix using one of three methods—\textit{full}, \textit{fragmented}, or \textit{lazy}—to produce a 1D vector representing the lead's signal amplitude at each step.}
    \label{fig:CutImage}
\end{figure*}

After text removal, the next step is to isolate the signal trace. This process begins with the image undergoing a thresholding process, which converts it into a black-and-white image using Otsu's binarization method \cite{otsu_threshold_1979}. This technique automatically calculates an optimal threshold to distinguish the foreground (signal) from the background (non-signal areas), based on pixel brightness variance across the image. 

Specifically, pixel brightness variance is quantified horizontally across rows to identify the positions of ECG lines, and vertically across columns to define the start and end points of each trace. These peaks in variance facilitate the precise and automated separation of individual traces, thereby defining their boundaries accurately. This method ensures that each trace of the ECG is clearly isolated for further processing.

\noindent\fbox{\begin{minipage}{\columnwidth}
\textbf{Input:} \textit{Paper-based ECG converted into an RGB matrix with the text pixels replaced by the average value of the surrounding pixels.}\\
\textbf{Output:} \textit{Binary matrix with 0 (black) representing the trace of the signal and (1) the background. Each trace can contain one or more leads (see~\cref{fig:CutImage}).}
\end{minipage}}

\subsubsection{Trace Extraction}
\label{sec:lead_extraction}
As illustrated in~\cref{fig:CutImage} A, the trace extraction phase involves segmenting the binarized matrix into distinct traces, where each trace contains one or more leads. The goal is to identify pixel coordinates that represent the leads and convert these into amplitude values at each time step. This process involves scanning the matrix from left to right using three distinct techniques, some of which is detailed below.

\textbf{Fragmented Extraction.}
The Fragmented Extraction technique, outlined in~\cref{algo:Fragmented}, refines the extraction process by specifically targeting the lowest-lit pixels in each column. In the context of ECG digitization, 'lit pixels' refer to those with a zero value, which indicates the presence of the ECG signal. It is designed to consider only the darkest group of pixels when gaps appear between clusters of lit pixels. This approach is particularly effective at ignoring extraneous markings, such as lead annotations, although it typically requires a longer processing time compared to other extraction methods. This method ensures that the data extracted from the ECG is as accurate and clean as possible, enhancing the quality of the digitization process.

\textbf{Full Extraction}. This technique involves calculating the average position of lit pixels in each column of the matrix. This method is efficient and generally preserves the signal amplitude well. However, it can occasionally be susceptible to inaccuracies caused by artifacts, such as text annotations (e.g., lead names) superimposed on the signal. These artifacts may skew the average position calculation, potentially affecting the fidelity of the extracted signal. Despite this, the Full Extraction method is favored for its speed and overall accuracy in environments where such artifacts are minimal or can be effectively managed. The pseudocode for this method is detailed in~\cref{algo:Full}. 

\textbf{Lazy Extraction.}
Lazy Extraction begins by establishing an 'anchor point' — defined as the average position of active pixels in the first column of the matrix. This anchor point serves as a reference for scanning the matrix from left to right. The method primarily checks for lit pixels at the anchor point's position in each subsequent column. If no lit pixels are found directly at the anchor point, the search expands above and below this point to locate the nearest active pixels. The pseudocode for this method is detailed in~\cref{algo:Lazy}.

This approach is notably fast and efficiently minimizes interference from artifacts such as superimposed text. However, its reliance on a fixed anchor point can occasionally lead to reduced signal fidelity, particularly when the ECG trace significantly deviates from the initial anchor position. This reduction in fidelity may impact the accuracy of the extracted signal, making it less suitable for applications requiring high precision in signal interpretation, such as detailed diagnostic analyses. Despite this, Lazy Extraction is advantageous in scenarios where speed is critical and minor inaccuracies are acceptable.

Noteworthy, each trace includes a reference pulse, typically found at the beginning of the signal (see~\cref{fig:CutImage}). This reference pulse, typically measuring 1 mV, serves as an amplitude scale to calibrate the rest of the signal. It is extracted simultaneously with the corresponding lead, ensuring that all amplitude measurements are accurately scaled.

\begin{algorithm}[h!]
\caption{Fragmented Extraction}
\label{algo:Fragmented}
\KwIn{Binarized matrix $M$ of size $n \times m$}
\KwOut{$V$ vector of size $m\times 1$ containing the lit-pixels coordinates.}
$V\gets \textit{zeros}((m, ))$\\
\For{$j \gets 0$ \KwTo $m$}{
    $\textit{lit\_pixels} \gets \text{lit\_pixels\_positions(M[:, j])}$\\
    $\textit{group\_of\_pixels} \gets []$
    
    \If{len(lit\_pixels) == 0}{
        $continue $    
    }
    \Else{
    $it \gets lit\_pixels[-1]$\\
    $l \gets -1$\\
    \While{it == lit\_pixels[l] }{
       $\textit{group\_of\_pixels}.append (lit\_pixels[l]) $ \\
       $it \gets it-1$\\
       $l \gets l-1$
       }
    $V[j] \gets mean(\textit{group\_of\_pixels})$ 
    }
    
}
\Return $V$
\end{algorithm}

\begin{algorithm}[h!]
\caption{Full Extraction}
\label{algo:Full}
\KwIn{Binarized matrix $M$ of size $n \times m$}
\KwOut{$V$ vector of size $m\times 1$ containing the lit-pixels coordinates.}

$V\gets \textit{zeros}((m, ))$\\
\For{$j \gets 0$ \KwTo $m$}{
    $\textit{lit\_pixels} \gets \text{lit\_pixels\_positions(M[:, j])}$\\
    \If{len(lit\_pixels) == 0}{
        $continue$
        
    }
    \Else{
         $V[j] \gets mean(lit\_pixels)$
    }
}
\Return $V$
\end{algorithm}

\begin{algorithm}[h!]
\caption{Lazy Extraction}
\label{algo:Lazy}
\KwIn{Binarized matrix $M$ of size $n \times m$}
\KwOut{$V$ vector of size $m\times 1$ containing the lit-pixels coordinates.}
$V\gets \textit{zeros}((m, ))$\\
$\textit{lit\_pixels} \gets \text{lit\_pixels\_positions}(M[:, 0])$

\If{$len(\textit{lit\_pixels}) == 0 $}{
    $a\_point \gets n/2$
}
\Else{
    $a\_point \gets \textit{mean}(\textit{lit\_pixels}(M[:, 0])$)
}
$V[0] \gets a\_point$ 

\For{$j \gets 1$ \KwTo $m$}{
    \If{$\text{lit\_pixel\_exists}(M[a\_point, j])$}{
        $V[j] \gets a\_point$  
    }
    \Else{
        $new\_a\_point \gets a\_point$\\
        \For{$i \gets 0$ \KwTo $n$ }{
            \If{$(a\_point +i \leq n)~and~
            lit\_pixel\_exists(M[a\_point+i, j])$}{
                 $new\_a\_point \gets a\_point+i$\\
                 \textbf{break}\
            }
            \ElseIf{$(a\_point - i \geq 0)~and~lit\_pixel\_exists(M[a\_point-i, j])$}{
                 $new\_a\_point \gets a\_point-i$\\ 
                 \textbf{break}\
            }
        }
        $a\_point \gets new\_a\_point$ \\
        $V[j] \gets a\_point$ \\
    }
}
\Return $V$

\end{algorithm}

\textbf{Lead Separation.}
After identifying the lit pixels, lead separation is achieved based on key assumptions: \textit{(1)} The first element in the vector containing the lit pixel coordinates marks the start of the reference pulse, while the last element indicates the end of the final lead. \textit{(2)} The ECG to be digitized typically adheres to one of two common standard formats: either a reference pulse followed by four leads in a 3x4 configuration, each lasting 2.5 seconds, or a reference pulse followed by two leads in a 2x6 configuration, each lasting 5 seconds. These formats are prevalent in clinical settings, making them the primary focus of our digitization efforts. While there are many other ECG formats, specific to particular recording devices, these will not be covered in this document due to their specialized nature and lesser prevalence. 

This methodological framework allows {\ecgtizer} to effectively standardize and automate the digitization process, ensuring reliable and consistent extraction of ECG data across the most commonly encountered formats.

The lead-separation procedure involves the following steps:
\begin{enumerate}
    \item Given the vector $V \subseteq [0:n]^{m \times 1}$, we \textbf{up-sample or down-sample it to 5,140 points}\footnote{Using linear interpolation (\texttt{numpy.interp}).}. This size is based on the assumption of 10-second recordings at a 500 Hz sampling rate, resulting in 5,000 points, with an additional 140 points specifically allocated for the reference pulse. This resized vector is referred to as $V_{\text{r}}$.
    \item Based on the assumptions above, the signal is segmented by slicing off the first 140 points to \textbf{isolate the reference pulse} ($R = V_{\text{r}}[0:140]$).
    \item The \textbf{leads are then isolated} by slicing every 1,250 points (for the 2.5-second format) or every 2,500 points (for the 5-second format). The vector representing a generic lead will be denoted as $L$.
    \item Once the leads and reference pulse have been isolated, the next step involves \textbf{scaling the extracted leads in terms of amplitude}. The reference pulse, serving as a calibration baseline, has its first pixels set to an amplitude of 0 mV, and its highest pixels to an amplitude of 1 mV. The final amplitude for each pixel in the lead is calculated using the equation:
    \begin{equation}
        L[i] = \frac{L[i] - \min(R)}{\max(R) - \min(R)},
    \end{equation}
    where $i$ ranges from 0 to 1,250 or 2,500, depending on the ECG format.
\end{enumerate}

\noindent\fbox{\begin{minipage}{\columnwidth}
\textbf{Input:} \textit{A binarized matrix where 0 (black) represents the trace of the ECG signal, and 1 (white) denotes the background. This matrix is derived from an initial processing step where the original grayscale or color image is converted into a binary format to simplify trace detection. Each trace within the matrix may contain one or more ECG leads.}

\textbf{Output:} \textit{After processing the digitized ECG image, each lead is represented as a one-dimensional vector. Each element in this vector corresponds to a specific point in time, with its value indicating the amplitude of the signal at that time step. The amplitude values are normalized based on reference points determined during the signal extraction process, typically ranging from 0 (minimum amplitude) to 1 (maximum amplitude), reflecting the relative intensity of the signal.}
\end{minipage}}

\subsubsection{Signal Completion}
In digitized ECGs, each lead is typically displayed over a specific time window. For example, in the \texttt{Standard} 3$\times$4 format, leads I, II, and III are shown for the first 2.5 seconds, followed by leads aVR, aVL, and aVF for the next 2.5 seconds, and so forth. Various methods have been developed to impute the missing portions of these leads \cite{seo_multiple_2022, joo2023twelve}. Among these, {\ecgrecover}~\cite{lence2024ecgrecoverdeeplearningapproach} is notable for its approach to reconstructing the missing sections of each lead by leveraging information from the available leads within the same time window. {\ecgrecover} utilizes a U-Net architecture, a type of deep learning model that is particularly effective in handling data with significant spatial and temporal dependencies. The model is trained to reconstruct missing ECG segments by minimizing a cost function designed to maintain the integrity and continuity of both the spatial alignment and temporal sequence of the lead data.

As the final step in the digitization process {\ecgrecover} allows users to generate a comprehensive 10-second recording that integrates all 12 leads. This synthesis not only fills in any missing data but also aligns and calibrates the leads to ensure the recording reflects a continuous and accurate representation of the cardiac activity.

\noindent\fbox{\begin{minipage}{\columnwidth}
\underline{Optional}

\textbf{Input:} \textit{Leads I, II, III, aVR, aVL, and aVF, each represented as a one-dimensional vector covering a 2.5-second timeframe. Each entry in the vector corresponds to a specific time point, with its value indicating the signal's amplitude at that moment.}

\textbf{Output:} \textit{The completed representation includes all 12 leads over a 10-second period, totaling 60,000 data points (12 leads x 5,000 points per lead). Each entry in this extended vector corresponds to a specific time point, with its value indicating the signal's amplitude. This comprehensive dataset enhances diagnostic capabilities by providing a full cardiac cycle representation for detailed analysis.}
\end{minipage}}

\subsection{Datasets}
\label{sec:experiments1}
In this work, we utilized three datasets: two for evaluating the digitizing methods (JOCOVID and PTB-XL) and one for downstream classification tasks (GENEREPOL).

\textbf{JOCOVID}. This dataset was acquired at the Pitié-Salpétrière hospital across various departments, including Cardiology, Internal Medicine, and Clinical Pharmacology during the COVID-19 pandemic~\cite{salem2022echocardiography}. The JOCOVID study (NCT04320017) is a retrospective observational study aimed at characterizing cardiac arrhythmias in COVID-19 hospitalized patients, conducted between March and May 2020. It consists of 807 10-second ECG recordings from 160 patients, available in both \texttt{.xml} and \texttt{.pdf} formats. The recordings, captured with an ELI 280 (Mortara device) at a sampling rate of 1,000 Hz with a 150 Hz filter under resting conditions, are displayed in a 3$\times$4 format with detailed segmentation of each lead. The cohort featured a male-to-female ratio of 1.6, a median age of 72 years [IQR=22], and a median follow-up duration of 12 days [IQR=7].

\textbf{PTB-XL}. The PTB-XL dataset~\cite{wagner2020ptb} is a publicly available collection of clinical ECGs recorded between 1989 and 1996, comprising 21,799 recordings from 18,869 patients. These 12-lead ECGs are each 10 seconds long, recorded at sampling frequencies of either 500 Hz or 300 Hz, and encompass a wide age range from 0 to 95 years, with a median of 62 years [IQR=22]. Approximately 44\% of the recordings are annotated as normal, while the remaining 56\% represent various pathologies, all annotated by specialist doctors. This dataset is widely used as a benchmark in ECG analysis research.

\textbf{GENEREPOL}. Extracted from the Generepol study~\cite{salem2017genome}, this dataset includes 15,119 10-second ECGs from 990 healthy volunteers, recorded at 500Hz with eight leads. Leads III, aVF, aVL, and aVR were computed from leads I, II, and V1-V6, following standard practices. The volunteers were part of a study conducted from 2008 to 2012 at the Clinical Investigation Center of Pitié-Salpétrière Hospital, where ECGs were recorded before and at several time points after administering 80mg of oral Sotalol. The recordings were curated by two expert cardiologists to ensure high data quality.

\subsection{Evaluation Metrics}
We assessed the performance of different methods in digitizing ECG signals by comparing the extracted numerical signals with the original signals stored in \texttt{.xml} files. For this evaluation, we utilized three main metrics: the Pearson Correlation Coefficient (PCC), the Root Mean Square Error (RMSE), and Soft-Distance Time Warping (SDTW). The PCC assesses the linear correlation between two signals, indicating their similarity in temporal dynamics. The RMSE measures the magnitude of differences between two signals, with smaller values indicating a closer match between the predicted and actual values. SDTW evaluates the alignment of two signals over time, accounting for potential time shifts. Additionally, we analyzed the accuracy of identifying specific ECG features (P, Q, R, S, T waves) using the \texttt{NeuroKit} Python library\footnote{\url{https://neuropsychology.github.io/NeuroKit/}}, which provides tools for clinically relevant feature detection in ECG signals. This comparison focused on the concordance of feature positioning between the extracted and original \texttt{.xml} signals as well as the derrived calculated time distances.

\subsection{State-of-the-art (SOTA) Comparison}
We compared the performance of two digitization tools, {\ecgminer}\footnote{\url{https://github.com/adofersan/ecg-miner}}~\cite{santamonica2024ecgminer}, a fully automated Python-based tool that optimizes pixel connectivity across regions of interest using a cost function, and {\paperecg}\footnote{\url{https://github.com/Tereshchenkolab/paper-ecg/tree/master}}\cite{Fortune2022}, which requires manual lead identification and uses Viterbi’s dynamic programming algorithm for signal extraction\cite{viterbi1967error}. Both tools were evaluated on the entire JOCOVID dataset of 807 ECGs. Additionally, we replicated the same evaluation on the PTB-XL dataset, but without including {\paperecg} as it required extensive human intervention. These results are presented in the supplementary materials (\cref{sup_mat}).

\subsection{Statistical Analyses}
Statistical analyses were conducted in R using linear mixed models, implemented through the \textit{lme4} and \textit{lmerTest} packages. The models included `patient id` and `lead` as random variables to account for intra-subject variability and differences among ECG leads, respectively. Figures and graphical representations of the data were generated using \textit{ggplot2} package and the whole \textit{tidyverse} suite, ensuring high-quality visualizations of our results.

\section{Results}
\label{sec:results1}
\begin{figure*}[htbp] 
    \centering
    \includegraphics[width=\textwidth]{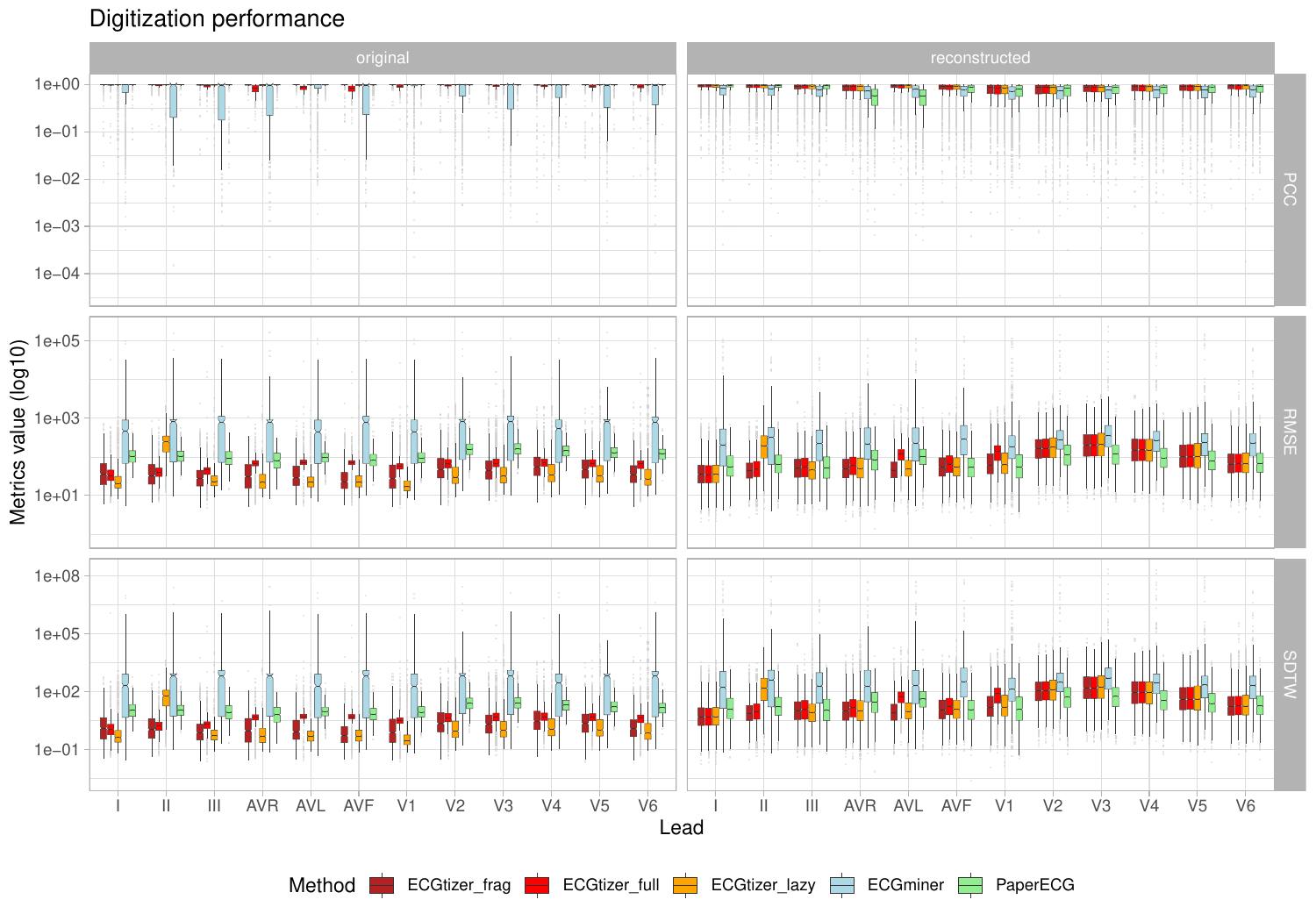} 
    \caption{Boxplot representation of digitization performance metrics (PCC, RMSE and SDTW) between the extracted signal and the real ECG signal before (original) and after the lead-completion phase (reconstructed). The evaluation of the reconstructed data is focused only on the recovered part of the signal.}
    \label{fig:PCC_RMSE_SDTW_plot}
\end{figure*}
In this section, we present the results of evaluating {\ecgtizer} for the digitization of ECG signals, with and without the optional lead-completion phase, as discussed in \cref{sec:discuss_distortion,sec:discuss_heartwaves} and detailed in \cref{sec:discussion_completion}.

\subsection{Digitization Performance}
\label{sec:results}
\begin{table}[htbp]
 \centering
 \scriptsize
 \caption{The table displays the digitization time for the JOCOVID dataset, along with the number of successfully digitized ECGs out of the total 807. The file is excluded from the final count if an error occurs during digitization. }
 \begin{tabular}{l | c |C{1.5cm}|C{1.5cm}|C{1.5cm}}
& Fully Automatic  & \centering Extracted ECG  & Time/ECG (seconds) & Time Total (hours)\\
\midrule

{\ecgtizer}$_{\text{Frag}}$ & Yes &  807 & 9 & 2\\
{\ecgtizer}$_{\text{Full}}$& Yes &  807 &  10 & 2.24 \\
{\ecgtizer}$_{\text{Lazy}}$ & Yes &  807 & 9 & 2 \\
{\ecgminer}  & Yes &  807 & 17 & 3.8\\
{\paperecg} &  No&  807 &  137 & 30.7\\ 
\end{tabular}
\label{tab:Performance_method_time}
\end{table}

\begin{figure*}[htbp]
    \centering
    \includegraphics[width=0.9\textwidth]{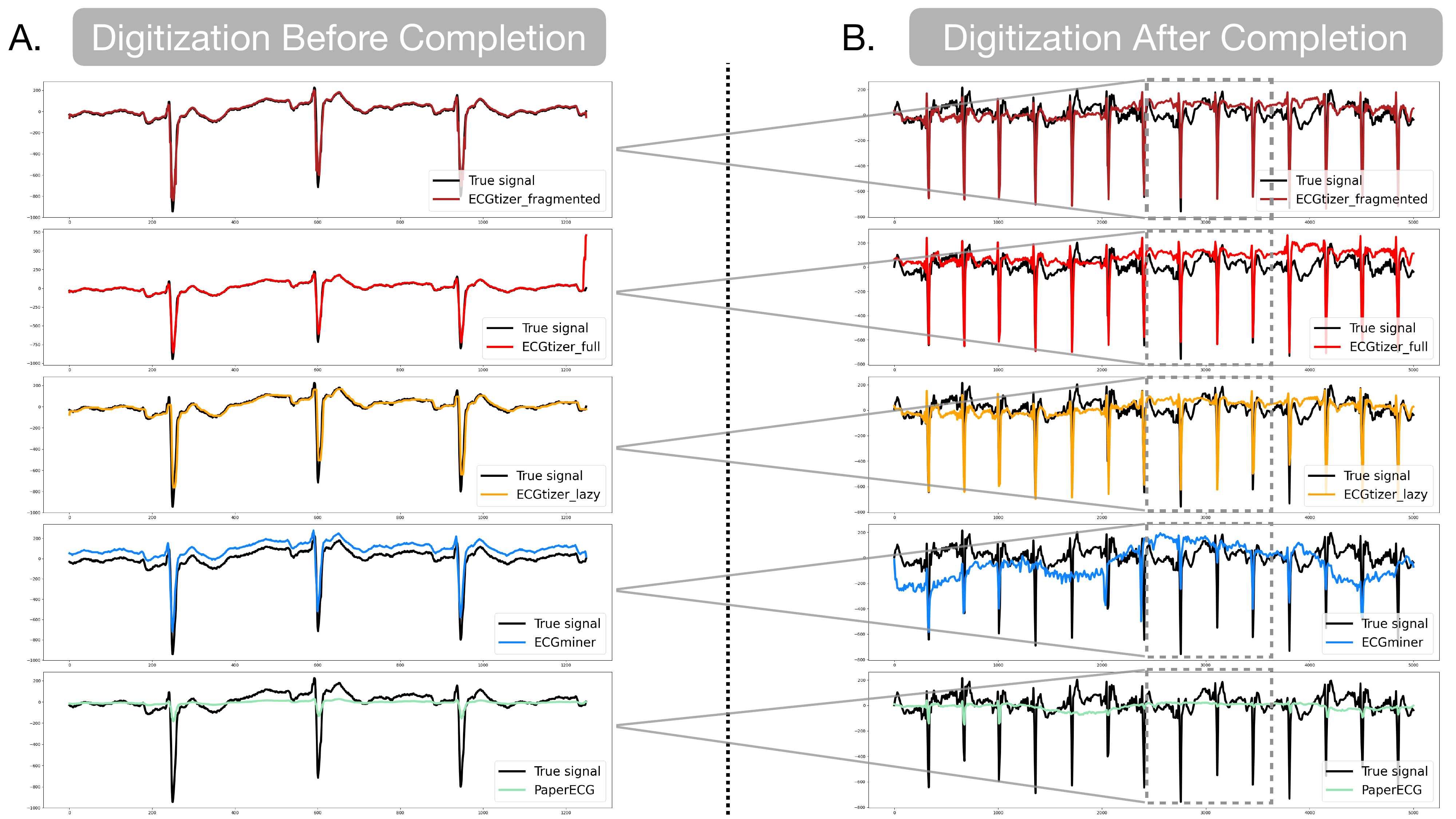}
    \caption{An illustration of the digitization output from the three methods {\ecgtizer}, {\ecgminer} and {\paperecg}, before and after completion with {\ecgrecover}. This example is extracted from the JOCOVID dataset and corresponds to the lead V1 of the patient n° 137.}
\label{fig:extraction}
\end{figure*}
We compare the performance of {\ecgtizer}, using three distinct methodologies, to {\ecgminer} and {\paperecg} across different evaluation metrics. These methodologies, referred to as {\ecgtizer}$_{\text{Full}}$, {\ecgtizer}$_{\text{Lazy}}$, and {\ecgtizer}$_{\text{Frag}}$ (described in \cref{sec:lead_extraction}), each employ a unique approach to signal extraction.

\subsubsection{Signal Fidelity}
\label{sec:discuss_distortion}
The results, summarized in \cref{fig:PCC_RMSE_SDTW_plot}, indicate that all three {\ecgtizer} methods outperform {\ecgminer} in terms of signal fidelity across all metrics. Specifically, {\paperecg} shows lower variability and comparable performance to the best {\ecgtizer} methods. 

In terms of the Pearson Correlation Coefficient (PCC), {\ecgtizer}$_{\text{Frag}}$ achieves an average correlation of 0.954, significantly higher than {\ecgminer}'s 0.672 (p<2e-16), but comparable to {\paperecg}'s 0.952 (p=0.51, not significant). {\ecgtizer}$_{\text{Full}}$ and {\ecgtizer}$_{\text{Lazy}}$ also perform well with correlations of 0.885 and 0.933, respectively.

For Root Mean Square Error (RMSE), which ideally should be as close to 0 as possible, {\ecgtizer} methods demonstrate significantly lower values than {\ecgminer} (902.4) and {\paperecg} (158.96, p<0.032). Specifically, RMSE values are 88.21 for {\ecgtizer}$_{\text{Frag}}$, 120.57 for {\ecgtizer}$_{\text{Full}}$, and 132.52 for {\ecgtizer}$_{\text{Lazy}}$.

Finally, Soft-Distance Time Warping (SDTW) scores, where lower values indicate better temporal alignment, further confirm the superiority of the {\ecgtizer} methods. {\ecgtizer}$_{\text{Frag}}$, {\ecgtizer}$_{\text{Full}}$, and {\ecgtizer}$_{\text{Lazy}}$ record scores of 916.8, 1388.5, and 1059.8, respectively, significantly outperforming {\ecgminer}'s 26432.5 (p<8.54e-09) and are similar to {\paperecg}'s 1443.1.

These results underscore the effectiveness of {\ecgtizer} in accurately digitizing ECG signals while maintaining high fidelity in signal reproduction compared to existing tools.

In \cref{fig:extraction}, we present an example of the digitized ECGs produced by {\ecgtizer}, {\ecgminer}, and {\paperecg}, compared with the original ECG signal (shown in black). The comparison is illustrated before (Panel A) and after (Panel B) the application of {\ecgrecover} to the digitized signals. This visual representation reveals two notable phenomena: first, both {\paperecg} and {\ecgminer} methods tend to underestimate the amplitude of the R peak, which could affect the clinical interpretation of the ECG. This is also the case for {\ecgtizer}$_{\text{Lazy}}$. Second, the amplitude scaling of {\paperecg} is not acurate, and for {\ecgminer} a notable misalignment is observed in the basal amplitude of the extracted ECGs compared with the true basal amplitude.

These discrepancies are critical as they could lead to misdiagnoses or errors in patient monitoring. The underestimation of the R peak amplitude, for instance, might affect the clinical assessments related to cardiac conditions that are often indicated by changes in this specific peak.

Finally, \Cref{tab:Performance_method_time} details the time required by each method to digitize an ECG from the Jocovid dataset. It is noteworthy that the {\paperecg} method requires approximately 15 times longer than the automated methods. More than 30 human hours were necessary to digitize 807 ECG with {\paperecg}, compared to 2 hours for {\ecgtizer} by a single unprallelized computer. Such differences in processing time are important considerations for clinical settings where rapid ECG analysis may be crucial.

\subsubsection{ECG Feature Detection}
\label{sec:discuss_heartwaves}
This section evaluates the ability of different digitizing tools to preserve crucial ECG features utilized by clinicians, specifically focusing on the P-wave, QRS complex, and T-wave, using lead II. This lead is commonly used to measure these features due to its clear depiction of the heart's electrical activity \cite{hampton2019ecg}. Using the \texttt{NeuroKit} Python library, each ECG heartbeat—defined as the segment centered around the R peak, with 200 data points before and 300 points after—was isolated as described in~\cite{golany2019pgans}. The digitized 10-second ECGs, scaled at 500 Hz, comprise a vector of 5,000 data points.

We analyzed time intervals, such as the QT interval and the QRS complex, along with the wave peak amplitudes. The QT interval is crucial for assessing the risk of arrhythmias like Torsades-de-Pointes (TdP), while the QRS complex duration and amplitude are indicators of conditions such as hypertrophic cardiomyopathy and pulmonary diseases \cite{power2021electrocardiographic}. Peak amplitudes, measured in millivolts (mV), provide insights into cardiac function and signal integrity.

The differences in QT and QRS durations and peak amplitudes between the digitized and original signals are summarized in \cref{tab:Performance_method}. We computed the absolute differences in these measurements to assess the fidelity of the digitization, with smaller differences indicating more accurate reproduction.

\Cref{fig:PCC_RMSE_SDTW_comp} illustrates the squared differences between the estimated features in the digitized and the original signals, displayed on a logarithmic scale to enhance clarity. Comparative statistics were performed using linear mixed models, with `Patient id' and `Features' as random variables, providing a robust framework for interpretation. The `Features' indicate the different measurements of positions and the resulting durations (QT, QRS).

In the time domain, {\paperecg} showed the smallest average absolute difference at 25.47 ms, closely followed by {\ecgtizer}$_{\text{Full}}$ at 28.02 ms, a difference which was not statistically significant (p=0.08). {\ecgtizer}$_{\text{Frag}}$, {\ecgminer}, and {\ecgtizer}$_{\text{Lazy}}$ showed larger differences, with {\ecgminer} and {\ecgtizer}$_{\text{Full}}$ differing significantly (p<6.15e-12).

In the amplitude domain, {\ecgtizer}$_{\text{Full}}$ and {\ecgtizer}$_{\text{Frag}}$ were equivalent, with average absolute differences of 74.79 $\mu V$ and 124.58 $\mu V$, respectively. They outperformed {\paperecg}, {\ecgtizer}$_{\text{Lazy}}$, and {\ecgminer}, which exhibited larger discrepancies in amplitude measurements, respectively 217.33, 345.88 and 895.41 $\mu V$ of average absolute difference. 

These findings highlight the effectiveness of {\ecgtizer} in accurately preserving essential ECG features, which is critical for reliable clinical assessments and research.

\begin{table*}[t]
    \centering
    \scriptsize
    \caption{$\Delta$ QT and $\Delta$ QRS represent the differences in segment lengths (in seconds), while $\Delta$ X$_{\text{amp}}$ represents the difference in peak amplitude (in mV), between the digitized and the original signal (before the lead-completion phase).}

\begin{tabular}{l|c|c|c|c|c}
& {\ecgtizer}$_{\text{Frag}}$ & {\ecgtizer}$_{\text{Full}}$ & {\ecgtizer}$_{\text{Lazy}}$ & {\ecgminer} & {\paperecg} \\
\midrule
$\Delta$P$_{\text{pos}}$         & \textbf{0.02 ± 0.06} & \textbf{0.02 ± 0.07} & 0.04 ± 0.11 & 0.03 ± 0.06 & \textbf{0.02 ± 0.05} \\
$\Delta$Q$_{\text{pos}}$& 0.02 ± 0.04 & 0.02 ± 0.03 & 0.03 ± 0.05 & \textbf{0.01 ± 0.03 }& \textbf{0.01 ± 0.03} \\
$\Delta$S$_{\text{pos}}$& \textbf{0.01 ± 0.03} &0.02 ± 0.04 & 0.04 ± 0.07 & 0.02 ± 0.05 & \textbf{0.01 ± 0.03}\\
$\Delta$T$_{\text{pos}}$& \textbf{0.03 ± 0.08} & 0.05 ± 0.1 & 0.08 ± 0.16 & 0.06 ± 0.09 & \textbf{0.03 ± 0.08}\\
$\Delta$QT          & 0.06±0.10 & \textbf{0.04±0.08} & 0.09±0.16 & 0.06±0.09 & \textbf{0.04±0.08} \\
$\Delta$QRS         & 0.03±0.05 & 0.03±0.05 & 0.06±0.08 & 0.03±0.05 & \textbf{0.02±0.04} \\
\midrule
\midrule
$\Delta$R$_{\text{amp}}$  & \textbf{0.11±0.18} & 0.13±0.20 & 0.85±0.74 & 0.49±2.05 & 0.55±0.34 \\
$\Delta$P$_{\text{amp}}$  & 0.04±0.08 & \textbf{0.03±0.08} & 0.12±0.14 & 0.93±6.13 & 0.07±0.07 \\
$\Delta$Q$_{\text{amp}}$  & 0.16±0.21 & \textbf{0.06±0.11} & 0.19±0.19 & 0.96±6.42 & 0.12±0.10 \\
$\Delta$S$_{\text{amp}}$  & 0.21±0.25 & \textbf{0.08±0.14} & 0.33±0.34 & 1.11±7.91 & 0.19±0.16 \\
$\Delta$T$_{\text{amp}}$  & 0.07±0.14 & \textbf{0.03±0.08} & 0.22±0.24 & 0.94±6.71 & 0.11±0.09 \\

\end{tabular}
\label{tab:Performance_method}
\end{table*}

\subsubsection{Recovering the missing signal}
\label{sec:discussion_completion}

At this step of the {\ecgtizer} pipeline, the ECG signal is only partial. In this section, we discuss the final and optional step in the {\ecgtizer} pipeline: lead completion. This process is essential for downstream analyses, such as advanced classification methods. To address this, we employ {\ecgrecover}~\cite{lence2024ecgrecoverdeeplearningapproach}, a deep learning model specifically designed to reconstruct missing parts of the ECG. Initially trained on the Generepol dataset~\cite{salem2017genome}, {\ecgrecover} was applied to digitized ECGs processed by the five methods previously discussed. This analysis, besides providing an evaluation of the digitizers in their capacity to recover the missing signal, allows to generate complete ECG for downstream analyses.

\Cref{fig:extraction} B illustrates an example of an ECG post-digitization and completion. The highlighted box indicates the segment originally digitized from the paper ECG (panel A). The reconstructed signal, while slightly different from the original, completes the entire ECG, offering a potential replacement for the digitized portion with a reconstructed equivalent.

\paragraph{Signal Fidelity after Signal Completion with {\ecgrecover}}
The effectiveness of the signal completion was quantitatively assessed by comparing the reconstructed signals to the original ECGs from the \texttt{.xml} files, focusing on metrics such as the Pearson Correlation Coefficient (PCC), Root Mean Square Error (RMSE), and Soft-Distance Time Warping (SDTW). The results, presented in~\cref{fig:PCC_RMSE_SDTW_plot} \textit{right}, highlight the accuracy of {\ecgrecover} restoring the full ECG waveform. This completion only enhances the usability of historical ECG data in downstream applications. 

\begin{figure*}[htbp]
    \centering
    \includegraphics[width=0.9\textwidth]{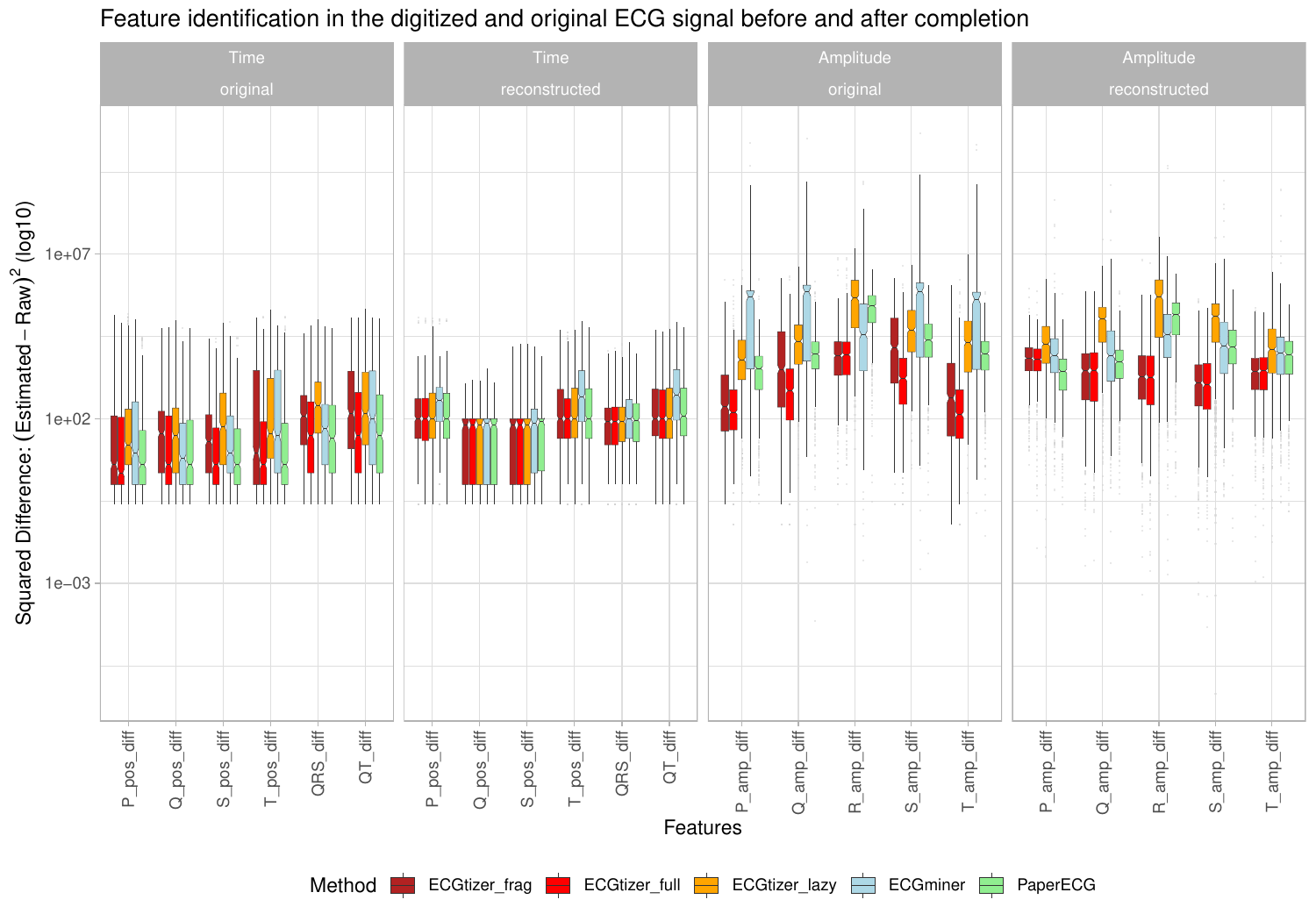}
    \caption{Boxplot representation of the squared differences between the identified features in the digitized signal and those identified in the raw original signal. Some features are estimated at the time level (left panel) and others (right panel) at the amplitude level both before (original) and after (reconstructed) signal recovery. These differences are shown in a logarithmic scale.}
\label{fig:PCC_RMSE_SDTW_comp}
\end{figure*}

When evaluating the reconstructed signals, {\ecgtizer} methods demonstrate notable efficacy. Specifically, {\ecgtizer}$_{\text{Frag}}$ exhibits the highest Pearson Correlation Coefficient (PCC) at 0.806, closely followed by {\ecgtizer}$_{\text{Full}}$ with a PCC of 0.800 and {\ecgtizer}$_{\text{Lazy}}$ at 0.798. The differences between {\ecgtizer}$_{\text{Frag}}$ and {\ecgtizer}$_{\text{Full}}$ are not statistically significant (p=0.11), indicating comparable performance in terms of correlation. Meanwhile, {\paperecg} and {\ecgminer} show lower correlations of 0.705 and 0.633, respectively.

In terms of Root Mean Square Error (RMSE), {\ecgtizer}$_{\text{Frag}}$, {\ecgtizer}$_{\text{Full}}$, and {\ecgtizer}$_{\text{Lazy}}$ achieve 129.89, 139.67, and 166.15, respectively, indicating precise signal reconstruction. {\paperecg} performs comparably well with an RMSE of 107.87, not significantly different from the best performing {\ecgtizer} methods. Conversely, {\ecgminer} lags significantly behind with an RMSE of 606.59 (p<2e-16).

For Soft-Distance Time Warping (SDTW), the three {\ecgtizer} methods again perform best, showing the lowest values: 201.75, 221.29, and 389.05 for {\ecgtizer}$_{\text{Frag}}$, {\ecgtizer}$_{\text{Full}}$, and {\ecgtizer}$_{\text{Lazy}}$ respectively. This contrasts starkly with {\ecgminer}, which records an exceedingly high SDTW of 88683.23, significantly worse than all other methods (p<2e-16). {\paperecg} achieves the best SDTW performance at 95.95, indicating excellent temporal alignment.

\begin{table*}[t]
    \centering
    \scriptsize
    \caption{$\Delta$ QT and $\Delta$ QRS represent the differences in segment lengths (in seconds), while $\Delta$ X$_{\text{amp}}$ represents the difference in peak amplitude (in mV) between the digitized and the original signal (after lead-completion phase with {\ecgrecover}).}

\begin{tabular}{l|c|c|c|c|c}
& {\ecgtizer}$_{\text{Frag}}$ & {\ecgtizer}$_{\text{Full}}$ & {\ecgtizer}$_{\text{Lazy}}$ & {\ecgminer} & {\paperecg} \\
\midrule
$\Delta$P$_{\text{pos}}$ & \textbf{0.03 ± 0.05} & \textbf{0.03 ± 0.05} &  \textbf{0.03 ± 0.06} & 0.04 ± 0.05 & \textbf{0.03 ± 0.05} \\
$\Delta$Q$_{\text{pos}}$ &\textbf{ 0.01 ± 0.01} & \textbf{0.01 ± 0.01} &  \textbf{0.01 ± 0.01} & \textbf{0.01 ± 0.02} & \textbf{0.01 ± 0.01} \\
$\Delta$S$_{\text{pos}}$ &\textbf{ 0.01 ± 0.02} & \textbf{0.01 ± 0.02} &  \textbf{0.01 ± 0.02} & 0.02 ± 0.03 & 0.02 ± 0.02 \\
$\Delta$T$_{\text{pos}}$ & \textbf{0.04 ± 0.06} & \textbf{0.04 ± 0.06} &  \textbf{0.04 ± 0.06} & 0.07 ± 0.07 & \textbf{0.04 ± 0.06} \\
$\Delta$QT          & \textbf{0.04±0.06} & \textbf{0.04±0.06} & \textbf{0.04±0.06} & 0.07±0.08 & \textbf{0.04±0.06} \\
$\Delta$QRS         & \textbf{0.02±0.03} & \textbf{0.02±0.03} & \textbf{0.02±0.03} & 0.03±0.04 & \textbf{0.02±0.03} \\
\midrule
\midrule
$\Delta$R$_{\text{amp}}$ & \textbf{0.08 ± 0.13} & \textbf{0.08 ± 0.13} & 0.85 ± 0.78 & 0.41 ± 0.29 & 0.41 ± 0.29 \\
$\Delta$P$_{\text{amp}}$ & 0.09 ± 0.07 & 0.09 ± 0.06 & 0.18 ± 0.19 & 0.18 ± 0.88 & \textbf{0.06 ± 0.05} \\
$\Delta$Q$_{\text{amp}}$ & \textbf{0.07 ± 0.09} & 0.08 ± 0.09 & 0.35 ± 0.31 & 0.24 ± 1.44 & 0.08 ± 0.07 \\
$\Delta$S$_{\text{amp}}$ & \textbf{0.06 ± 0.08} & \textbf{0.06 ± 0.07} & 0.41 ± 0.38 & 0.31 ± 1.72 & 0.17 ± 0.14 \\
$\Delta$T$_{\text{amp}}$ & \textbf{0.06 ± 0.06} & 0.07 ± 0.06 & 0.18 ± 0.23 & 0.19 ± 1.15 & 0.11 ± 0.08 \\

\end{tabular}
\label{tab:Performance_method_comp}
\end{table*}

Overall, the results indicate that the methods proposed by {\ecgtizer} and {\paperecg} lay a solid foundation for subsequent reconstruction with {\ecgrecover}. Across all evaluated metrics—PCC, RMSE, and SDTW—the {\ecgtizer}$_{\text{Fragmented}}$ technique consistently outperforms the other {\ecgtizer} variants. Moreover, for the PCC metric, the methods proposed by {\ecgtizer} surpass current state-of-the-art techniques, showcasing their robustness in maintaining temporal dynamics of the ECG signal.

While {\paperecg} exhibits slightly better results than {\ecgtizer}$_{\text{Frag}}$ for RMSE and SDTW metrics, the differences are not statistically significant. This suggests that both methods are comparably effective at minimizing errors and aligning the signals over time, despite slight variations in their operational algorithms.

Interestingly, there is a noted decrease in performance after the signal completion phase. This decline is anticipated as the completion process inherently modifies the original signal. Specifically, it tends to attenuate noise, which can make the completed ECG appear less similar to the original, particularly at higher frequencies. However, this modification is crucial for reducing artifacts and preserving essential clinical information that is critical for accurate diagnosis and analysis.

\paragraph{ECG Feature Identification Fidelity After Signal Completion}

In the assessment of feature detection fidelity in the time domain for the reconstructed signal, {\ecgtizer}$_{\text{Frag}}$ demonstrates the best performance with a minimal average absolute time difference of 25.9 ms, closely followed by {\ecgtizer}$_{\text{Full}}$ at 26.27 ms, with no significant difference between them. {\ecgtizer}$_{\text{Lazy}}$, {\paperecg}, and {\ecgminer} follow, with time differences of 27.85 ms, 28.21 ms, and 41.37 ms, respectively. In the amplitude domain, {\ecgtizer}$_{\text{Frag}}$ and {\ecgtizer}$_{\text{Full}}$ also lead, both exhibiting an average absolute difference of around 80 $\mu$V, significantly outperforming the other methods. These results are visualized in~\cref{fig:PCC_RMSE_SDTW_comp}.

\Cref{tab:Performance_method_comp} indicates that {\ecgrecover} partially corrects the digitized ECG, notably improving QT and QRS measurements. This improvement might be attributed to {\ecgrecover}'s ability to enhance peak detection by clarifying the signal, thus facilitating the isolation of peak-related pixels. Another contributing factor could be the initial measurement approach, which was based on averaging only 2 to 3 beats. After completion, averaging extends across 10 beats, likely reducing the impact of minor errors.

It is important to highlight that while {\ecgrecover} is primarily designed for reconstructing missing segments of the ECG, it appears to also clean the signal, although this was not an intended function during its training, as discussed in the reference paper~\cite{lence2024ecgrecoverdeeplearningapproach}. Despite these enhancements in time domain measurements, amplitude measurements see no improvement post-completion, reflecting {\ecgrecover}'s design focus, which does not include modifying signal amplitude.

In the subsequent section, we will delve deeper into the characteristics of the ECG post-completion to further analyze how these modifications impact the clinical utility of the reconstructed signals. This analysis will help us understand the trade-offs involved in enhancing signal fidelity while maintaining clinical relevance.

\subsection{Downstream Classification Tasks}
\label{sec:tdp}

As we conclude our evaluation of {\ecgtizer}, we transition from analyzing its performance in signal digitization and feature reconstruction to assessing its utility in a critical clinical application. Specifically, we estimate its effectiveness in classifying the IKr-blockade ECG footprint, a crucial marker for predicting the risk of Torsades-de-Pointes. This classification task, as demonstrated by Prifti et al.~\cite{prifti_deep_2021}, relies heavily on the accurate detection and representation of specific ECG features that {\ecgtizer} has been shown to preserve effectively.

The ability of {\ecgtizer} to maintain high fidelity in the reconstructed signals suggests that it could significantly enhance the reliability of downstream classification tasks in cardiology. By ensuring that crucial waveforms and intervals are accurately captured and represented, {\ecgtizer} supports advanced analytical methods that predict potentially life-threatening arrhythmias like Torsades-de-Pointes.

For this experiment, we utilized the Generepol dataset (refer to \cref{sec:methodology}), focusing exclusively on ECGs recorded before (Sot-) and 3 hours after (Sot+) Sotalol intake, which are critical for assessing TdP risk. The dataset comprised 3,066 Sot- and 2,734 Sot+ ECGs. Notably, while the original ECGs were in {\texttt{.xml}} format, we converted them to a paper-like format using the {\texttt{ecg2pdf}} library\footnote{\url{https://github.com/RigacciOrg/ecg-contec/tree/main}}, adapting the process to ensure compatibility with our digitizing tools. However, due to format incompatibility, the ECGs generated were not readable by {\ecgminer}, necessitating an alternative conversion method as described in~\cite{santamonica2024ecgminer}. Moreover, {\paperecg} was excluded from this experiment due to the extensive time required for manual lead annotation.

We employed a CNN model, originally developed by Prifti et al.~\cite{prifti_deep_2021}, to predict TdP risk. This DenseNet-style model features six blocks, each with eight dense convolutional layers, and was trained over 200 epochs using the Adam optimizer (learning rate: 0.001, dropout rate: 0.2). We trained five versions of the model: one on the original {\texttt{.xml}} data and four on data digitized via {\ecgtizer} and {\ecgminer}. Given that ECG lead II is inherently 10 seconds long in the standard 4$\times$4 paper format, this lead required no further completion or reconstruction, focusing our analysis solely on this lead.

In \cref{tab:ModelsOriginal_perf}, we compare the performance of the model trained on the original data with that on the digitized testing set from the Generepol dataset. A noticeable performance decline was observed across the digitized data compared to the original. Notably, the most significant declines were seen with {\ecgtizer}$_{\text{Lazy}}$ and {\ecgminer}, which also showed the greatest discrepancies in peak amplitude (\cref{sec:discuss_heartwaves}), underscoring the importance of peak amplitude accuracy for TdP-risk prediction. Interestingly, the original model's relatively robust performance suggests that accurate QT distance detection, a feature well-preserved in the original data, may play a critical role in effective TdP-risk prediction.

\begin{table}[htbp]
    \centering
    \scriptsize
    \caption{
    TdP-risk prediction models trained on the original data in \texttt{.xml} format from Generepol and tested on digitized Generepol data and on the original testing set. }
    \begin{tabular}{p{1.95cm}| c | c | c | c  }
& Accuracy & Precision & Recall & F1-Score   \\
\midrule

{\ecgtizer}$_\text{Frag}$ & 0.83 & \textbf{0.93} & 0.70 & 0.80  \\
{\ecgtizer}$_\text{Full}$       & \textbf{0.85} & 0.91 & \textbf{0.75} & \textbf{0.82}  \\
{\ecgtizer}$_\text{Lazy}$       & 0.76 & 0.69 & 0.82 & 0.75    \\
{\ecgminer}       & 0.47 & 0.47 & 1.00 & 0.64    \\
\midrule
Original data  & 0.94 & 0.95 & 0.92 & 0.93 \\            
\end{tabular}
\label{tab:ModelsOriginal_perf}
\end{table}
\begin{table}[htbp]
    \centering
    \scriptsize
    \caption{
    TdP-risk prediction models trained and tested on the original data (\texttt{.xml} format) from Generepol or on the corresponding digitized data from Generepol.}
    \begin{tabular}{p{1.95cm} | c | c | c | c  }
& Accuracy & Precision & Recall & F1-Score   \\
\midrule
{\ecgtizer}$_\text{Frag}$ & 0.88 & \textbf{0.89} & 0.85 & 0.86  \\
{\ecgtizer}$_\text{Full}$       & \textbf{0.89} & \textbf{0.89} & \textbf{0.86} & \textbf{0.87}  \\
{\ecgtizer}$_\text{Lazy}$       & 0.80 & 0.88 & 0.65 & 0.75    \\
{\ecgminer}       & 0.79 & 0.79 & 0.74 & 0.76    \\
\midrule
Original data       & 0.94 & 0.95 & 0.92 & 0.93 \\            
\end{tabular}
\label{tab:Models_perf}
\end{table}

In \cref{tab:Models_perf}, we present the performance of the DenseNet model retrained on digitized training data from the same dataset with the same framework. This retraining led to improved performance compared to the results in \cref{tab:ModelsOriginal_perf}, where models were tested on digitized data without retraining. However, despite these gains, the performance levels with digitized data still did not match those achieved with the original {\texttt{.xml}} data. Among the {\ecgtizer} methods, {\ecgtizer}$_{\text{Full}}$ showed the best results, underscoring its capacity to preserve essential features critical for accurate TdP-risk prediction.

One plausible explanation for the persistent performance gap can be linked to the inherent loss of information during the paper conversion. As outlined in another study by Granese et al.~\cite{granese2023negative}, basic alterations to the signal, particularly through processes like cleaning and pre-processing, can detrimentally impact deep learning models. In addition, the digitization process itself includes operations such as smoothing and resampling, which might also contribute to subtle but clinically significant alterations, thus negatively affecting the performance of the DenseNet model. These changes can obscure or alter key ECG features that deep learning algorithms, like DenseNet, rely on for making accurate predictions.

This phenomenon highlights a critical challenge in applying machine learning to clinical diagnostics: maintaining the integrity of original data features during preprocessing and digitization is essential for preserving model accuracy. Moving forward, it is crucial to refine these methods to minimize information loss and better simulate the characteristics of original ECG recordings, ensuring that advanced predictive models can perform optimally in clinical settings. Future research could further refine {\ecgtizer}'s algorithms to optimize for specific classification tasks or explore its integration into broader clinical workflows. The ongoing development and validation of {\ecgtizer} in these practical applications hold the potential to improve diagnostic accuracy and patient outcomes in cardiac care.

\section{Conclusion}
\label{sec:conclusion}

In this paper, we introduced {\ecgtizer}, an innovative open-source tool designed to automatically digitize paper-based ECGs in the Standard 4x4 clinical format. A unique feature of {\ecgtizer} is its ability to reconstruct missing segments of the ECG, ensuring complete 10-second coverage for each of the 12 leads, a capability that sets it apart from other state-of-the-art methods.

{\ecgtizer} has shown superior performance in our evaluations, particularly with its three signal extraction algorithms: {\ecgtizer}${\text{Full}}$, {\ecgtizer}${\text{Frag}}$, and {\ecgtizer}${\text{Lazy}}$. Notably, {\ecgtizer}${\text{Frag}}$ excels in preserving the QT/QRS segments critical for accurate ECG interpretation, while {\ecgtizer}$_{\text{Full}}$ has demonstrated particular effectiveness in supporting downstream AI tasks sensitive to specific ECG features such as peak amplitudes.

Despite these promising results, our study has limitations, primarily focusing on digitally generated, clean paper-like ECGs from original XML files. In real-world applications, paper ECGs often present additional challenges, including annotations, creases, and variable scan qualities. Future work will involve extensive testing of {\ecgtizer} under these more complex conditions to confirm its effectiveness in typical clinical environments.

Looking forward, we aim to enhance {\ecgtizer}’s adaptability to a variety of ECG formats used by single-lead recording devices. This development will broaden its applicability across different clinical tools and healthcare systems, potentially transforming ECG digitization practices for better diagnostic and therapeutic outcomes.

In conclusion, {\ecgtizer} constitutes a substantial innovation in the field of electrocardiography digitization. As an open-source and fully automated tool, it is uniquely positioned to enhance the scope and speed of electrocardiology research by facilitating access to and utilization of extensive historical datasets from well-characterized cohorts globally.


\section*{Acknowledgements}
This study was supported by the ANR-20-CE17-0022 DeepECG4U funding from the French National Research Agency (ANR).

\section*{Declaration of generative AI and AI-assisted technologies in the writing process}
During the preparation of this work the author(s) used ChatGPT 4 in order to improve the readibility of the text. After using this tool/service, the author(s) reviewed and edited the content as needed and take(s) full responsibility for the content of the published article.

\bibliographystyle{cas-model2-names}
\bibliography{main}

\appendix
\clearpage
\section{Additional Results on PTB-XL database}
\label{sup_mat}
In this section, we present the results of applying {\ecgtizer} to a public database, the PTB-XL database describe in \cref{sec:experiments1} . This database contains ECG recordings in digital format, which first required a step to convert this digital format into images, as discussed in section \cref{sec:tdp}.

For this analysis, we have chosen to focus on \textit{signal fidelity }and \textit{ECG feature detection}.

The main aim of this section is to demonstrate {\ecgtizer}'s ability to adapt to other databases. In addition, a sample of the PTB-XL database and instructions for converting it to PDF format are available on the {\ecgtizer} Git repository.

\subsection{Signal fidelity}
\paragraph{Before Completion}

\begin{figure*}[htbp] 
    \centering
    \includegraphics[width=\textwidth]{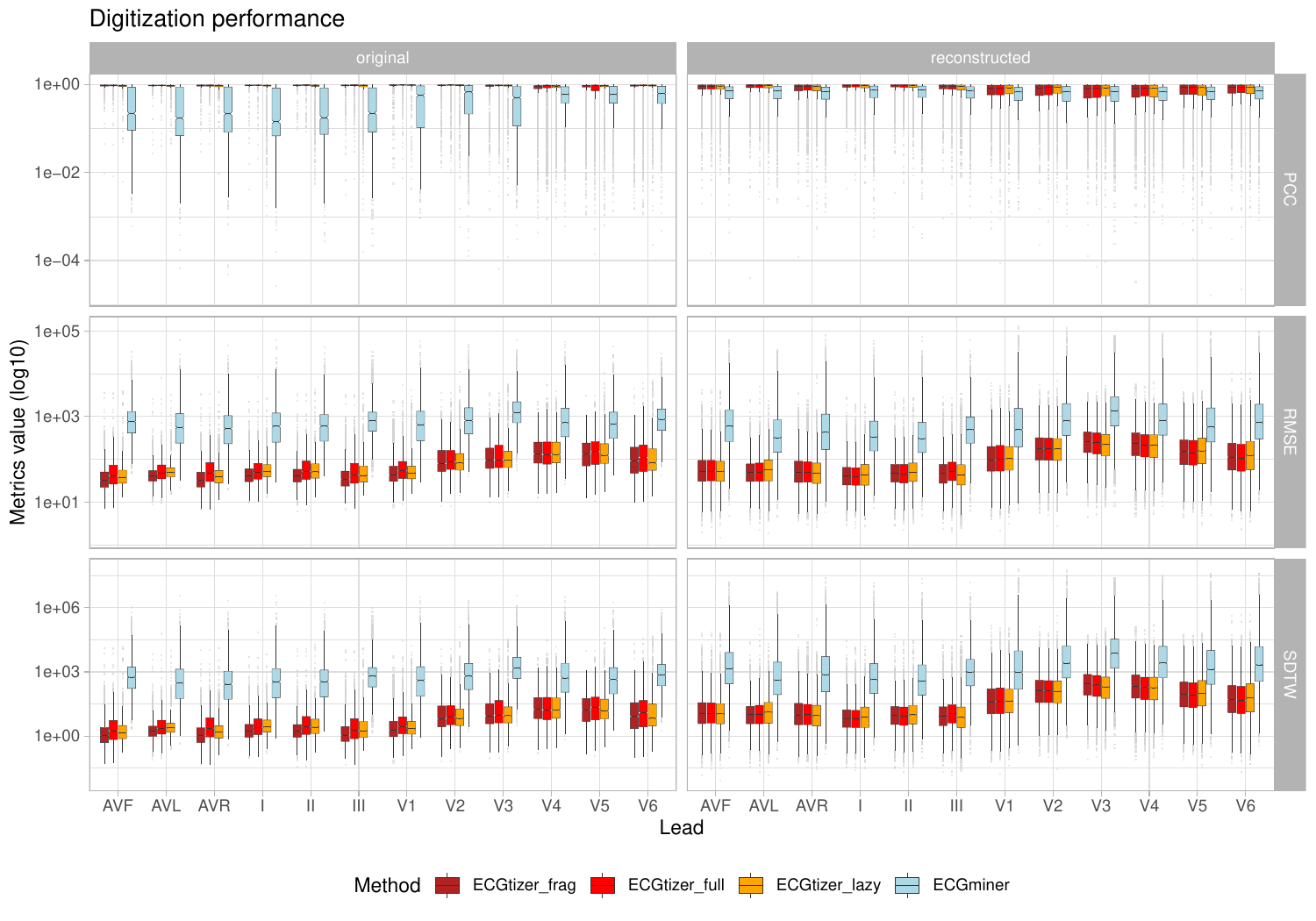} 
    \caption{Boxplot representation of digitization performance metrics (PCC, RMSE and SDTW) between the extracted signal and the real ECG signal before (original) and after the lead-completion phase (reconstructed). The evaluation of the reconstructed data is focused only on the recovered part of the signal.}
    \label{fig:PCC_RMSE_SDTW_PTB_XL}
\end{figure*}

Overall in \cref{fig:PCC_RMSE_SDTW_PTB_XL} (original), we can see that the methods proposed by {\ecgtizer} significantly outperform the {\ecgminer} approach for all the metrics evaluated, namely PCC, RMSE and SDTW. PCC scores reach 0.861, 0.87 and 0.86 respectively for {\ecgtizer}${\text{Fragmented}}$, {\ecgtizer}${\text{Full}}$ and {\ecgtizer}$_{\text{Lazy}}$, while {\ecgminer} only achieves an PCC of 0.37. The {\ecgtizer}${\text{Full}}$ version appears significantly superior to the {\ecgtizer}${\text{Fragmented}}$ version ($p = 0.01$) for this metric.

In terms of RMSE, the {\ecgtizer}${\text{Fragmented}}$ and {\ecgtizer}${\text{Full}}$ perform best, with values of 103.79 and 121.43 respectively. Conversely, the {\ecgtizer}$_{\text{Lazy}}$ and {\ecgminer} show significantly lower results (p = 0.0001), with RMSE values of 149.48 and 1397.82 respectively.

With regard to SDTW, the different {\ecgtizer} methods show no significant differences, with scores ranging from 35.52 to 168.75. In contrast, {\ecgminer} displays a significantly lower score ($p < 2e-16$), reaching a value of 9077.85.

\paragraph{After Completion}

We completed the signal in accordance with the methods described in \cref{sec:discussion_completion}. The results, calculated only on the missing portions of the signal, are presented in \cref{fig:PCC_RMSE_SDTW_PTB_XL} (reconstructed).

Overall, the results show that the various methods proposed by {\ecgtizer} outperform those proposed by {\ecgminer}. 
Among the {\ecgtizer} approaches, the Fragmented and Full versions stand out, without showing significant differences after reconstruction on all the metrics analyzed (PCC, RMSE and SDTW). With a PCC of around 0.76, RMSE of 165 and 157 and SDTW of 357 and 300.

The {\ecgtizer}$_{\text{Lazy}}$ is slightly inferior to the other {\ecgtizer} variants in terms of PCC alone. In terms of RMSE and STDW, this method equals {\ecgtizer}${\text{Fragmented}}$ and {\ecgtizer}${\text{Full}}$.

As for {\ecgminer}, it performs significantly worse on all metrics after signal completion, with a PCC of 0.57, an RMSE of 1732 and an SDTW of 99857.

\subsection{ECG feature detection}
\paragraph{Before Completion}

\begin{figure*}[htbp]
    \centering
    \includegraphics[width=\textwidth]{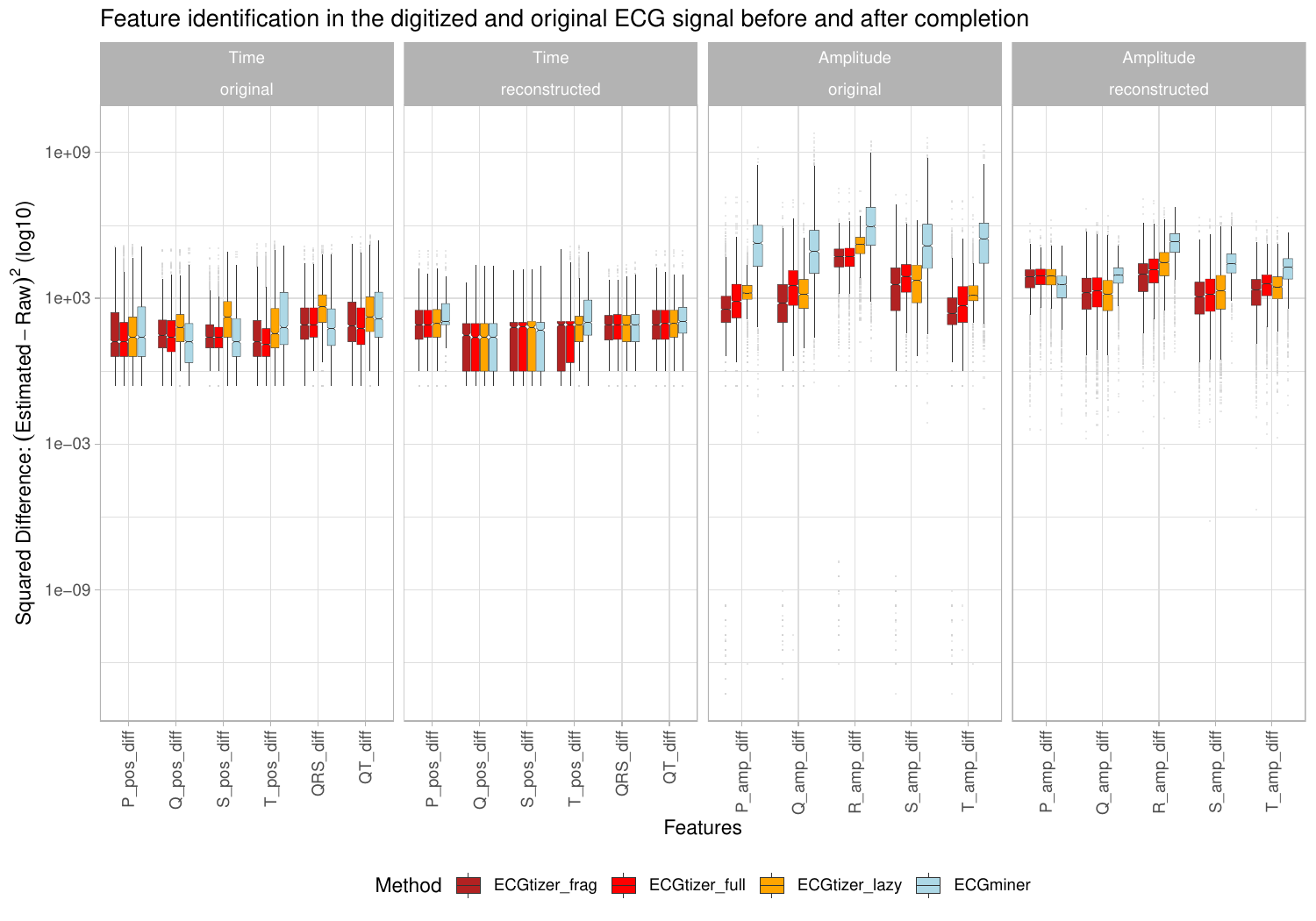}
    \caption{Boxplot the squared differences between the identified features in the digitized signal and those identified in the raw original signal. Some features are estimated at the time level (left panel) and others (right panel) at the amplitude level. These differences are shown in a logarithmic scale.}
\label{fig:Peaks_PTB_XL}
\end{figure*}
\begin{table*}[t]
    \centering
    \scriptsize
    \caption{$\Delta$ QT and $\Delta$ QRS represent the differences in segment lengths (in seconds), while $\Delta$ X$_{\text{amp}}$ represents the difference in peak amplitude (in mV) between the digitized and the original signal (on PTB-XL database).}
\begin{tabular}{l|c|c|c|c}
& {\ecgtizer}$_{\text{Frag}}$ & {\ecgtizer}$_{\text{Full}}$ & {\ecgtizer}$_{\text{Lazy}}$ & {\ecgminer} \\
\midrule
$\Delta$P$_{\text{pos}}$ & \textbf{0.04 ± 0.09} & \textbf{0.04 ± 0.1} & 0.05 ± 0.11 & \textbf{0.04 ± 0.08} \\
$\Delta$Q$_{\text{pos}}$ & \textbf{0.02 ± 0.04} & \textbf{0.02 ± 0.04} & 0.03 ± 0.06 & 0.02 ± 0.04 \\
$\Delta$S$_{\text{pos}}$ & \textbf{0.02 ± 0.04} & \textbf{0.02 ± 0.03} & 0.04 ± 0.06 & 0.03 ± 0.06 \\
$\Delta$T$_{\text{pos}}$ & \textbf{0.04 ± 0.08} & 0.03 ± 0.08 & 0.07 ± 0.14 & 0.06 ± 0.1 \\
$\Delta$QT          & \textbf{0.05 ± 0.09} & \textbf{0.05 ± 0.09} & 0.08 ± 0.15 & 0.07 ± 0.11 \\
$\Delta$QRS         & \textbf{0.03 ± 0.05} & 0.06 ± 0.08 & 0.04 ± 0.06 & 0.06 ± 0.06 \\
\midrule
\midrule
$\Delta$R$_{\text{amp}}$ & \textbf{0.24 ± 0.23} & 0.27 ± 0.25 & 0.44 ± 0.28 & 2.1 ± 3.83 \\
$\Delta$P$_{\text{amp}}$ & \textbf{0.04 ± 0.13} & 0.05 ± 0.13 & 0.05 ± 0.1 & 0.88 ± 1.76 \\
$\Delta$Q$_{\text{amp}}$ & \textbf{0.06 ± 0.15} & 0.09 ± 0.15 & \textbf{0.06 ± 0.11} & 1.14 ± 3.87 \\
$\Delta$S$_{\text{amp}}$ & \textbf{0.11 ± 0.18} & 0.13 ± 0.17 & \textbf{0.11 ± 0.14} & 1.31 ± 3.49 \\
$\Delta$T$_{\text{amp}}$ & \textbf{0.04 ± 0.14} & 0.05 ± 0.13 & 0.06 ± 0.12 & 1.02 ± 2.19 \\

\end{tabular}
\label{tab:Performance_method_ptb_xl}
\end{table*}
Looking at the \cref{tab:Performance_method_ptb_xl}, we can clearly see that the {\ecgtizer} set of methods outperforms {\ecgminer}. Furthermore, the \cref{fig:Peaks_PTB_XL}(original) shows that the {\ecgtizer}${\text{Full}}$ and {\ecgtizer}${\text{Fragmented}}$ methods achieve the best results.

Concerning the coherence in feature detection in the time domain for the original digitization, the first place is taken by {\ecgtizer}$_{\text{Full}}$ with an average absolute distance of 32.42 ms, followed by {\ecgtizer}$_{\text{Fragmented}}$ with an average absolute distance of 32.48 ms, however, this difference is not significant (p=0.95) and they can be considered as equivalent. They are followed by {\ecgminer} with 41.86 ms and {\ecgtizer}$_{\text{Lazy}}$ with 55.84 ms. The difference between {\ecgtizer}$_{\text{Fragmented}}$ and {\ecgminer} is statistically significant ($p<2e-16$).

For the coherence in feature detection in the amplitude domain in the original signal, the first place is taken by {\ecgtizer}$_{\text{Frag}}$, which is equivalent to {\ecgtizer}$_{\text{Full}}$ with a respective average absolute difference of 98.423 and 118,125 $\mu V$. They are followed by  {\ecgtizer}$_{\text{Lazy}}$ and {\ecgminer} with respectively 146,422 and 1288,758 $\mu V$ of average absolute 

\paragraph{After Completion}
\begin{table*}[t]
    \centering
    \scriptsize
    \caption{$\Delta$ QT and $\Delta$ QRS represent the differences in segment lengths (in seconds), while $\Delta$ X$_{\text{amp}}$ represents the difference in peak amplitude (in mV) between the digitized reconstructed signal and the original one (on PTB-XL database).}





\begin{tabular}{l|c|c|c|c}
& {\ecgtizer}$_{\text{Frag}}$ & {\ecgtizer}$_{\text{Full}}$ & {\ecgtizer}$_{\text{Lazy}}$ & {\ecgminer} \\
\midrule
$\Delta$P$_{\text{pos}}$          & 0\textbf{.02 ± 0.03} & \textbf{0.02 ± 0.04}
 & \textbf{0.02 ± 0.03} & 0.03 ± 0.04 \\
$\Delta$Q$_{\text{pos}}$          & \textbf{0.01 ± 0.01} & \textbf{0.01 ± 0.01} & \textbf{0.01 ± 0.01} & \textbf{0.01 ± 0.02} \\
$\Delta$S$_{\text{pos}}$         & \textbf{0.01 ± 0.02 }& \textbf{0.01 ± 0.02} & \textbf{0.01 ± 0.02} & \textbf{0.01 ± 0.02} \\
$\Delta$T$_{\text{pos}}$          & 0.03 ± 0.06 & \textbf{0.02 ± 0.06} & 0.03 ± 0.07 & 0.04 ± 0.06 \\
$\Delta$QT          & \textbf{0.02 ± 0.03} & \textbf{0.02 ± 0.04} & \textbf{0.02 ± 0.03} & 0.03 ± 0.03 \\
$\Delta$QRS         & \textbf{0.02 ± 0.02} & \textbf{0.02 ± 0.02} & \textbf{0.02 ± 0.02} & \textbf{0.02 ± 0.03} \\
\midrule
\midrule
$\Delta$R$_{\text{amp}}$ & \textbf{0.12 ± 0.14} & 0.17 ± 0.18 & 0.21 ± 0.18 & 1.45 ± 2.16 \\
$\Delta$P$_{\text{amp}}$ & \textbf{0.09 ± 0.07} & 0.1 ± 0.06 & 0.1 ± 0.07 & 0.42 ± 1.61 \\
$\Delta$Q$_{\text{amp}}$ & 0.07 ± 0.1 & 0.07 ± 0.08 & \textbf{0.06 ± 0.09} & 0.98 ± 2.73 \\
$\Delta$S$_{\text{amp}}$ & \textbf{0.06 ± 0.08} & 0.07 ± 0.09 & 0.07 ± 0.1 & 1.16 ± 3.1 \\
$\Delta$T$_{\text{amp}}$ & \textbf{0.06 ± 0.07} & 0.08 ± 0.07 & 0.07 ± 0.08 & 0.4 ± 1.05 \\

\end{tabular}
\label{tab:Performance_method_ptb_xl}
\end{table*}

Results for the absolute difference between extracted and real ECG are shown in \cref{tab:Performance_method_ptb_xl}, while the absolute difference for certain key features is shown in the \cref{fig:Peaks_PTB_XL}(reconstructed).

A reduction in error on temporal features is observed after completion. The {\ecgtizer}${\text{Frag}}$ approach achieves the best performance, followed by {\ecgtizer}${\text{Full}}$, with absolute errors of 18.81 and 19.14 ms respectively. The {\ecgtizer}$_{\text{Lazy}}$ and {\ecgminer} methods performed less well, with absolute errors of 20.00 and 24.61 ms respectively ($p < 0.05$).

In terms of amplitude-domain characteristics, {\ecgtizer}${\text{Frag}}$ achieved the best results, with an error of 79.06 µV. In comparison, the other methods perform significantly worse ($p < 2e10-16$), with errors of 94.24 µV, 103.31 µV and 205.21 µV respectively for {\ecgtizer}${\text{Full}}$, {\ecgtizer}$_{\text{Lazy}}$ and {\ecgminer}.

\clearpage
\appendix
\end{document}